\documentclass[prd,twocolumn,nopacs,floatfix,amsmath,nofootinbib,amssymb,floatfix]{revtex4}
\usepackage{graphicx,color,dcolumn,booktabs,bm}
\usepackage{longtable,lscape}
\usepackage{txfonts}
\usepackage{overpic}
\usepackage{enumerate}
\usepackage{indentfirst}
\usepackage{feynmf,slashed}   
\usepackage{cases}
\usepackage{color}
\usepackage{multirow}
\usepackage{epstopdf}
\usepackage{ulem}
\usepackage[colorlinks, citecolor=blue,anchorcolor=red,menucolor=red, linkcolor=red,filecolor=red,urlcolor=blue,frenchlinks=red]{hyperref}

\graphicspath{{Figures/}} %
\begin{document}

\title{Predicted $\Lambda\bar{\Lambda}$ and $\Xi^-\bar{\Xi}^+$ decay modes of the charmoniumlike $Y(4230)$}
\author{Ri-Qing Qian$^{1,2}$}\email{qianrq18@lzu.edu.cn}
\author{Qi Huang$^4$}\email{huangqi@ucas.ac.cn}
\author{Xiang Liu$^{1,2,3}$\footnote{Corresponding author}}\email{xiangliu@lzu.edu.cn}

\affiliation{$^1$School of Physical Science and Technology, Lanzhou University, Lanzhou 730000, China\\
$^2$Research Center for Hadron and CSR Physics, Lanzhou University $\&$ Institute of Modern Physics of CAS, Lanzhou 730000, China\\
$^3$Lanzhou Center for Theoretical Physics, Key Laboratory of Theoretical Physics of Gansu Province,
and Frontiers Science Center for Rare Isotopes, Lanzhou University, Lanzhou 730000, China\\
$^4$School of Physical Sciences, University of Chinese Academy of Sciences (UCAS), Beijing 100049, China}
\date{\today}

\begin{abstract}
In this work, we predict the light hadronic decay channels $Y(4230)\to\Lambda\bar{\Lambda}$ and $\Xi^-\bar{\Xi}^+$ when treating the $Y(4230)$ as a vector charmonium state. By the hadronic loop mechanism, the branching ratios of the $Y(4230)\to\Lambda\bar{\Lambda}$ and $\Xi^-\bar{\Xi}^+$ processes are calculated. In addition, we discuss the possibility of carrying out the search for the signal of the $Y(4230)$ through $\Lambda\bar{\Lambda}$ and $\Xi^-\bar{\Xi}^+$ channels
from the $e^+e^-$ annihilation. Assuming $Y(4230)$ exist in $\Lambda\bar{\Lambda}$ and $\Xi^-\bar{\Xi}^+$ channel, we also present the time-like electromagnetic form factors (EMFFs) at $\sqrt{s}=m_{Y(4230)}$. 
\end{abstract}

\maketitle

\section{\label{sec1}Introduction}
Although more and more charmonia have been observed since 1974 \cite{Aubert:1974js,Augustin:1974xw},
it is not the end of whole story of constructing the charmonium family, especially with a series of observations of charmoniumlike $XYZ$ states \cite{Liu:2019zoy,Brambilla:2019esw}. As an important group, the charmonium family by adding more states may provide crucial hints to understand the nonperturbative behavior of strong interaction, which has a close relation to the color confinement in Quantum Chromodynamics (QCD).

The vector charmoniumlike $Y$ state around 4.2 GeV plays a special role in constructing the charmonium family. In 2014, the Lanzhou group predicted the existence of a vector charmonium $\psi(4S)$ by borrowing the similarity between $J/\psi$ and $\Upsilon$ families \cite{He:2014xna}. In Ref.~\cite{He:2014xna}, the mass of the predicted $\psi(4S)$ is 4263 MeV, which is consistent with former prediction of $\psi(4S)$ by the screening potential model \cite{Li:2009zu}. Obviously, this study proposed a new task for the future experiments at that time. As indicated in Ref. \cite{Yuan:2013uta}, a possible enhancement structure around 4.2 GeV may exist in the experimental data of the $e^+e^-\to h_c\pi^+\pi^-$ \cite{BESIII:2013ouc} and the $e^+e^-\to J/\psi \pi^+\pi^-$ \cite{Belle:2013yex} processes, which should be checked by more precise experimental studies. In 2015, the BESIII Collaboration analyzed the $e^+e^-\to \chi_{c0}\omega$ process \cite{Ablikim:2014qwy}, where an obvious enhancement structure with mass $M=4230\pm8$ MeV and width $\Gamma=38\pm12$ MeV was found. Focusing on this novel phenomenon, the Lanzhou group indicated that this enhancement around 4.2 GeV in the $\chi_{c0}\omega$ invariant mass spectrum can be understood through introducing the predicted $\psi(4S)$ \cite{Chen:2014sra}. Also, in Ref.~\cite{Chen:2015bma}, the Lanzhou group found the evidence of $\psi(4S)$ in the $e^+e^-\to \psi(2S)\pi^+\pi^-$ data reported by the Belle Collaboration \cite{Belle:2014wyt}, where the extracted resonance parameters are $M=4243\pm7$ MeV and $\Gamma=16\pm 31$ MeV. Additionally, by performing a combined fit to the $e^+e^-\to \psi(2S)\pi^+\pi^-$ \cite{Belle:2014wyt}, $h_c\pi^+\pi^-$ \cite{BESIII:2013ouc} and $\chi_{c0}\omega$ \cite{BESIII:2014rja} processes, similar resonance parameters were obtained for the narrow structure around 4.2 GeV~\cite{Chen:2015bma}. In 2017, the BESIII Collaboration studied the $e^+e^-\to \psi(2S)\pi^+\pi^-$ process based on 5.1 fb$^{-1}$ of data \cite{BESIII:2017tqk} and confirmed the existence of the charmoniumlike structure around 4.2 GeV in the $\psi(2S)\pi^+\pi^-$ invariant mass spectrum predicted by the Lanzhou group. Besides these studies of hidden-charm decays, BESIII found the evidence of a structure in the $e^+e^-\to \pi^+ D^0 D^{*-}$ reaction around 4.2 GeV \cite{BESIII:2018iea}.
Focusing on these phenomena of the vector charmoniumlike structure around 4.2 GeV, further study from the Lanzhou group indicated that this structure plays a role of scaling point when constructing the $J/\psi$ family \cite{Wang:2019mhs}, where the assignment of the mixture state of $\psi(4S)$ and $\psi(3D)$ to the vector charmoniumlike structure around 4.2 GeV was proposed to describe the present experimental data \cite{BESIII:2014rja,Belle:2014wyt,BESIII:2016bnd,BESIII:2016adj,BESIII:2018iea}.
Here, this vector charmoniumlike at 4.2 GeV is referred to be the $Y(4230)$~\footnote{In Particle Data Group (PDG) \cite{ParticleDataGroup:2020ssz}, the $Y(4230)$ is known as $\psi(4230)$. Also, the former $Y(4260)$~\cite{BaBar:2005hhc} was omitted from PDG since the $Y(4230)$ was reported as substructure contained in the $Y(4260)$~\cite{BESIII:2016bnd}.}.
{Besides the charmonium assignment to the $Y(4230)$, there are other theoretical interpretations including a $D\bar{D}_1(2420)$ molecule~\cite{Cleven:2013mka,Ding:2008gr}, a hybrid~\cite{Zhu:2005hp,Close:2005iz,Kou:2005gt}, a compact tetraquark state~\cite{Maiani:2014aja}, a hadrocharmonium~\cite{Dubynskiy:2008mq,Li:2013ssa}. To better understand the nature of the $Y(4230)$, we should pay more efforts.}

\begin{figure}[htbp]
  \centering
  \includegraphics[width=8cm,keepaspectratio]{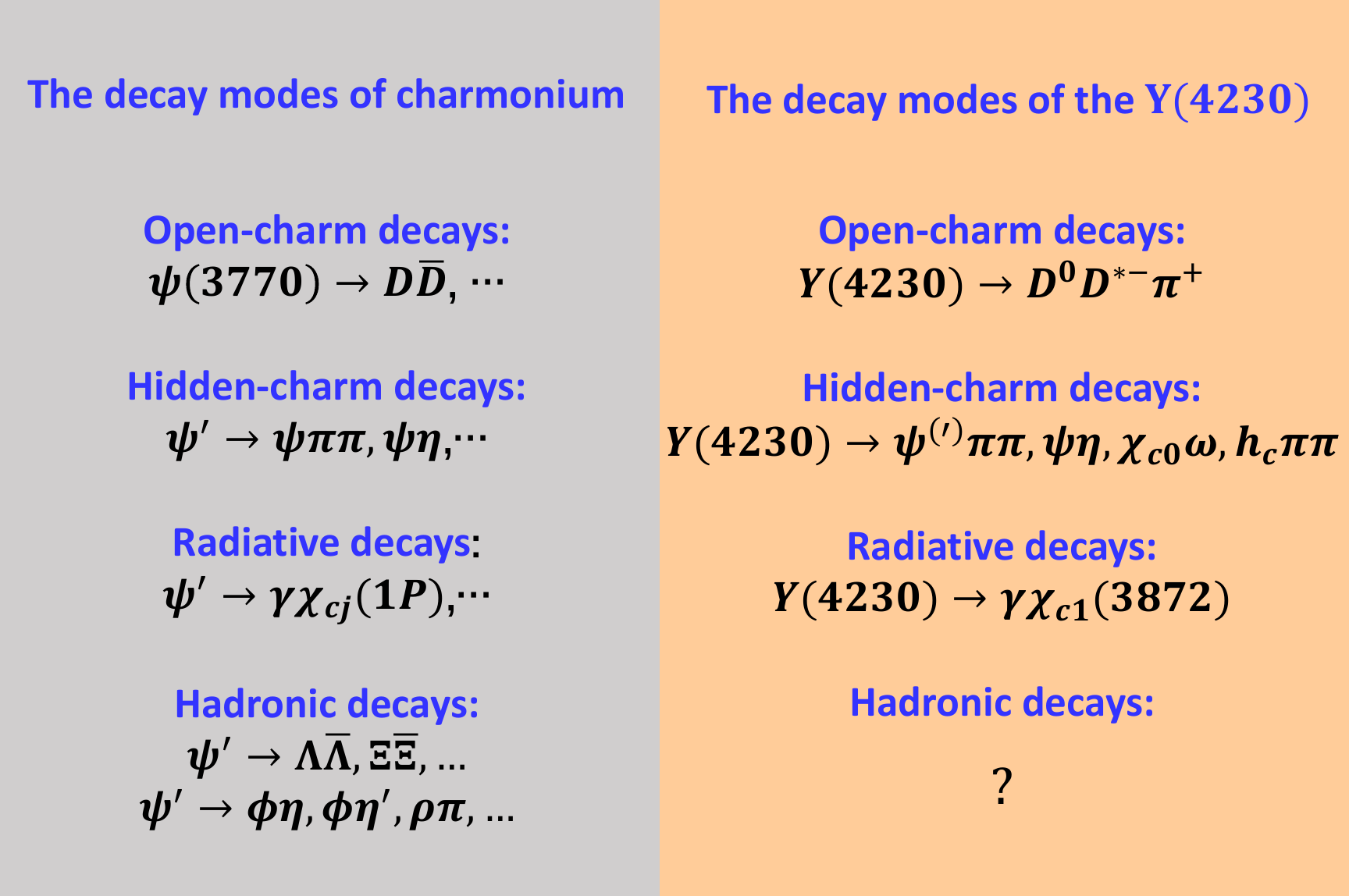}
  \caption{The typical decay modes of charmonium states and the discovered decay modes of the $Y(4230)$.}\label{fig:decay_modes}
\end{figure}

For charmonium states, there exist abundant decay modes if checking the PDG data \cite{ParticleDataGroup:2020ssz}. These modes, including the charmonium decays into hidden-charm channel, open-charm channel, light hadronic channels and so on. Here, we should specify the decay modes of charmonium into light hadrons.
As briefly reviewed above, for the $Y(4230)$, BESIII already found its hidden-charm \cite{BESIII:2014rja,Belle:2014wyt,BESIII:2016bnd,BESIII:2016adj}, radiative~\cite{BESIII:2019qvy} and possible open-charm decay modes~\cite{BESIII:2018iea} (see Fig.~\ref{fig:decay_modes}). At present, the $Y(4230)$ decays into light hadrons are still missing in experiment. If the $Y(4230)$ is a charmonium as suggested in Refs. \cite{Wang:2019mhs}, we have reason to believe that the $Y(4230)$ may decay into light hadrons. Searching for this kind of decay of the $Y(4230)$ will be an important task when identifying the properties of the $Y(4230)$.

Along this line, in this work we focus on the study of the $Y(4230)\to \Lambda\bar{\Lambda}$ and $\Xi^-\bar{\Xi}^+$ decays. Under the charmonium assignment to the $Y(4230)$, the coupled-channel effect on the $Y(4230)$ cannot be ignored, hence we introduce the hadronic loop mechanism in the concrete calculation of the branching ratios of the $Y(4230)\to \Lambda\bar{\Lambda}$ and $\Xi^-\bar{\Xi}^+$ decays.

We notice the measurements on the cross sections of the $e^+e^-\to \Lambda\bar\Lambda$ \cite{BESIII:2021ccp} and $e^+e^-\to \Xi^-\bar{\Xi}^+$ processes from the BESIII Collaboration \cite{BESIII:2019cuv}.
When checking the experimental data, we find possible event accumulation around 4.2 GeV in the $\Lambda\bar\Lambda$ and $\Xi^-\bar{\Xi}^+$ invariant mass spectrum. Thus, we discuss the possibility of searching for the $Y(4230)$ in the $\Lambda\bar{\Lambda}$ and $\Xi^-\bar{\Xi}^+$ decay channels by combining with the BESIII data.
We hope that the present study may stimulate the experimentalist's interest in establishing the existence of the $Y(4230)$ through the
$e^+e^-\to \Lambda\bar\Lambda$ and $e^+e^-\to \Xi^-\bar{\Xi}^+$ processes when more experimental data will be collected around
4.2 GeV energy point.

There is another reason for us to carry out this study. Recently, several measurements on the time-like electromagnetic form factors (EMFFs) of the hyperons have been done. For example, the EMFFs of $\Lambda$ were measured at $\sqrt{s}=m_{J/\psi}$~\cite{BESIII:2018cnd} and $\sqrt{s}=2.396$ GeV~\cite{BESIII:2019nep}, the EMFFs of $\Xi^-$ were obtained at $\sqrt{s}=m_{J/\psi}$~\cite{BESIII:2021ypr}, the EMFFs of $\Sigma^+$ were measured at $\sqrt{s}=m_{J/\psi}$ and $\sqrt{s}=m_{\psi^\prime}$~\cite{BESIII:2020fqg}. In these experiments, both module and phase of EMFFs were extracted via the subsequent decay of the hyperons. We should point out that at specific resonance point such as $\sqrt{s}=m_{J/\psi}$, these EMFFs actually provide a complete measurement of the amplitude of resonance decay, and the properties of the resonance can be inferred from the EMFFs. Hence, the hyperon anti-hyperon pairs serves as special decay modes of vector charmonium states. 

This paper is organized as follow.
After the introduction, we investigate the $\Lambda\bar{\Lambda}$ and $\Xi^-\bar{\Xi}^+$ decay modes of the $Y(4230)$ in Sec.~\ref{formalism}, where the hadronic loop mechanism introduced in these decay channels is introduced in Sec.~\ref{hadronicloop} and numerical results are present in Sec.~\ref{results}. In Sec.~\ref{data}, we discuss the searching possibility of the $Y(4230)$ in experiments, where the potential signals of the $Y(4230)$ is discussed. Based the hypothetical the $Y(4230)$ signals in the present data, the electromagnetic form factors of $\Lambda$ and $\Xi^-$ at $\sqrt{s}=m_{Y(4230)}$ is calculated in Sec.~\ref{formfactors}. Finally, this paper ends with a short summery.

\section{$Y(4230) \to \Lambda \bar{\Lambda}$ and $Y(4230)\to \Xi^-\bar{\Xi}^+$\label{formalism}}

\subsection{\label{hadronicloop}The hadronic loop mechanism in $Y(4230) \to \Lambda \bar{\Lambda}$ and $\Xi^-\bar{\Xi}^+$ porcesses}
In Ref.~\cite{Wang:2019mhs}, it was pointed out that the $Y(4230)$ can be a good candidate of charmonium state. Thus in this section, assigning the $Y(4230)$ to be a charmonium state, we investigate its decay into hyperon and antihyperon. Generally, these higher charmonia above the $D\bar{D}$ threshold have dominant open-charm decay channel, which contributes to their total decay widths. Since the coupled-channel effect may play important role to mediate the decays of higher charmonia, we should consider the hadronic loop contribution to the decay channels involved in light hadrons for these discussed higher charmonia.
Similar ideas have been successfully applied to the exploration of the hidden-charm and hidden-bottom decay of charmonia and bottomonia \cite{Liu:2006dq,Liu:2009dr,Zhang:2009kr,Meng:2007tk,Meng:2008dd,Meng:2008bq,Chen:2011qx,Chen:2011zv,Chen:2011pv,Chen:2014ccr,Wang:2016qmz,Huang:2017kkg,Huang:2018pmk,Huang:2018cco}.
{A typical example is that the prediction of the branching ratio $\mathcal{B}(\Upsilon(5S)\to\eta\Upsilon_J(1D))$ using the hadronic loop mechanism~\cite{Wang:2016qmz} was confirmed by the Belle experiment~\cite{Belle:2018hjt}. Especially, to $Y(4230)$ itself, in Refs. \cite{Chen:2014sra,Chen:2015bma,Cleven:2016qbn}, after introducing the hadronic loop mechanism involved with charmed mesons, its decay to $\chi_{c0}\omega$ and $\psi(2S)\pi^+\pi^-$ channels are studied and the consistency between theoretical and experimental results shows that $D$ meson loop plays very important roles in the hidden charm decays of the $Y(4230)$. Thus, applying the hadronic mechanism to explore the baryon pair decay mode of the $Y(4230)$ is straightforward. So}
in this work, we study the hadronic loop effect to the processes $Y(4230) \to \Lambda \bar{\Lambda}$ and $\Xi^-\bar{\Xi}^+$.

\begin{figure}[htbp]
  \centering
  \includegraphics[width=8cm,keepaspectratio]{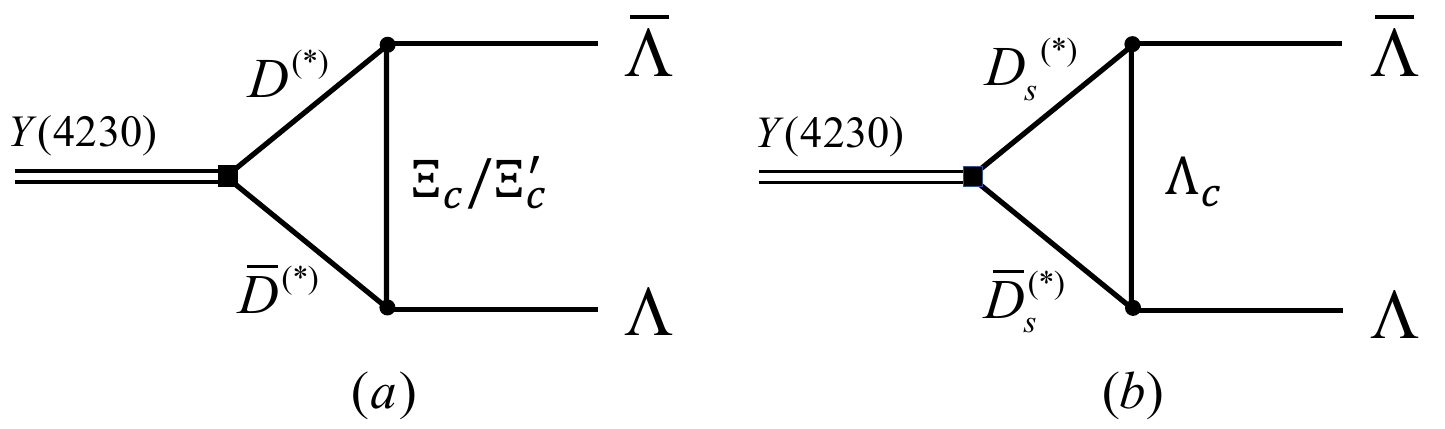}
  \caption{The schematic diagrams for the $Y(4230)\to\Lambda \bar{\Lambda}$ process of under hadronic loop mechanism.}\label{fig:lambda}
\end{figure}

Since the allowed open-charm decay channels of the $Y(4230)$ include $D\bar{D}$, $D\bar{D}^*+c.c.$, $D_s\bar{D}_s$ and $D_s\bar{D}_s^*+c.c.$, we should take into account the triangle loops composed of charmed mesons and charmed baryons, which is the bridge to connect the initial $Y(4230)$ and the $\Lambda\bar{\Lambda}$ final state as shown in Fig.~\ref{fig:lambda}. If the intermediate charmed meson pair is $D^{(*)}\bar{D}^{(*)}$, the exchanged charmed baryon can be either $\Xi_c$ or $\Xi_c^\prime$, while, if the intermediate meson pair is $D^{(*)}_{s}\bar{D}^{(*)}_s$, the exchanged charmed baryon is $\Lambda_c$.

To evaluate these processes given in Fig.~\ref{fig:lambda} at hadronic level, we first need the Lagrangians to describe the relevant interaction vertices. For the $Y(4230)D^{(*)}D^{(*)}$ interaction, we adopt the following form \cite{Huang:2018cco}

\begin{equation}\label{eq:lagrangianpsi}
   \begin{split}
     & \mathcal{L}_{\psi \mathcal{D}^{(*)}\mathcal{D}^{(*)}} \\
       & = -ig_{\psi \mathcal{DD}}\psi^\mu \left( \mathcal{D}^{\dagger}\overleftrightarrow{\partial}_\mu \mathcal{D} \right) \\
       & \quad +g_{\psi \mathcal{D}\mathcal{D}^*}\epsilon^{\mu\nu\alpha\beta}\partial_\mu\psi_\nu \left( \mathcal{D}^{\dagger} \overleftrightarrow{\partial}_\alpha \mathcal{D}^*_{\beta}-\mathcal{D}_\beta^{*\dagger}\overleftrightarrow{\partial}_\alpha \mathcal{D} \right)\\
       & \quad +ig_{\psi \mathcal{D}^*\mathcal{D}^*}\psi^\mu \left( \partial^\nu \mathcal{D}_\mu^{*\dagger} \mathcal{D}^*_\nu-\mathcal{D}_\nu^{*\dagger}\partial^\nu \mathcal{D}_\mu^*+\mathcal{D}^{*\nu\dagger}\overleftrightarrow{\partial}_\mu \mathcal{D}_\nu^* \right).
   \end{split}
 \end{equation}
For depicting the interaction of $B_c D^{(*)}\Lambda$, where $B_c$ denotes the charmed baryons, we use the following Lagrangian \cite{Khodjamirian:2011jp}
\begin{equation}\label{eq:lagrangianLa}
   \begin{split}
     \mathcal{L}_{B_cD^{(*)}_{(s)}\Lambda} &= ig_{B_cD_{(s)}\Lambda}\bar{B}_c\gamma^5D_{(s)}\Lambda \\
       & \quad +\bar{B}_c\left( g_{B_cD^*_{(s)}\Lambda}\gamma^\mu + \frac{\kappa_{B_cD^*_{(s)}\Lambda}}{m_{B_c}+m_\Lambda}\sigma^{\mu\nu}\partial_\nu \right) D^*_{(s)\mu} \Lambda +h.c..
   \end{split}
 \end{equation}

With the above preparation, we get the decay amplitudes of the $Y(4230)\to\Lambda\bar{\Lambda}$ process corresponding to these diagrams in Fig.~\ref{fig:lambda}, the amplitudes corresponding to different charmed meson combinations are
 \begin{equation}\label{ampbeg}
   \begin{split}
     \mathcal{M}^a & = i^3 \int \frac{d^4q_3}{(2\pi)^4} \left(-ig_{\psi DD}\right)\epsilon_\psi^\mu\left(iq_{2\mu}-iq_{1\mu}\right) \\
       & \quad \times \bar{u}(p_2) \left(ig_{B_cD\Lambda}\gamma^5\right)\left(\slashed{q}_3+m_{B_c}\right)\left(ig_{B_cD\Lambda}\gamma^5\right)v(p_1)\\
       & \quad \times \frac{1}{q_1^2-m_D^2} \frac{1}{q_2^2-m_D^2} \frac{1}{q_3^2-m_{B_c}^2} \mathcal{F}^2(q_3^2),
   \end{split}
 \end{equation}
 \begin{equation}
   \begin{split}
     \mathcal{M}^b & = i^3 \int \frac{d^4q_3}{(2\pi)^4} g_{\psi DD^*} \epsilon_\psi^\mu \epsilon_{\alpha\beta\lambda\mu} \left(iq_2^\alpha-iq_1^\alpha\right)\left(-ip^\lambda\right) \\
       & \quad \times \bar{u}(p_2)\left(ig_{B_cD\Lambda}\gamma^5\right)\left(\slashed{q}_3+m_{B_c}\right) \\
       & \quad \times \left(g_{B_cD^*\Lambda}\gamma^\rho-i\frac{\kappa_{B_cD^*\Lambda}}{m_{B_c}+m_\Lambda}\sigma^{\rho\nu}q_{2\nu}\right)v(p_1) \\
       & \quad \times \frac{1}{q_1^2-m_D^2} \frac{-g^\beta_\rho+q_2^\beta q_{2\rho}/m_{D^*}^2}{q_2^2-m_{D^*}^2} \frac{1}{q_3^2-m_{B_c}^2} \mathcal{F}^2(q_3^2),
   \end{split}
 \end{equation}
 \begin{equation}
   \begin{split}
     \mathcal{M}^c & = i^3 \int \frac{d^4q_3}{(2\pi)^4} g_{\psi DD^*} \epsilon_\psi^\mu \epsilon_{\alpha\beta\lambda\mu} \left(iq_1^\alpha-iq_2^\alpha\right)\left(-ip^\lambda\right) \\
       & \quad \times \bar{u}(p_2)\left(g_{B_cD^*\Lambda}\gamma^\rho-i\frac{\kappa_{B_cD^*\Lambda}}{m_{B_c}+m_\Lambda}\sigma^{\rho\nu}q_{1\nu}\right) \\
       & \quad \times \left(\slashed{q}_3+m_{B_c}\right) \left(ig_{B_cD\Lambda}\gamma^5\right)v(p_1) \\
       & \quad \times \frac{-g^\beta_\rho+q_2^\beta q_{2\rho}/m_{D^*}^2}{q_1^2-m_{D^*}^2} \frac{1}{q_2^2-m_{D}^2} \frac{1}{q_3^2-m_{B_c}^2} \mathcal{F}^2(q_3^2),
   \end{split}
 \end{equation}
 \begin{equation}\label{amped}
   \begin{split}
     \mathcal{M}^d & = i^3 \int \frac{d^4q_3}{(2\pi)^4} ig_{\psi D^*D^*}\epsilon_\psi^\mu \left[\left(iq_{2\mu}-iq_{1\mu}\right)g_{\alpha\nu}+iq_{1\alpha}g_{\mu\nu}\right.\\
       & \quad \left.-iq_{2\nu}g_{\mu\alpha}\right] \bar{u}(p_2)\left(g_{B_cD^*\Lambda}\gamma^\lambda-i\frac{\kappa_{B_cD^*\Lambda}}{m_{B_c}+m_{\Lambda}}\sigma^{\lambda\rho}q_{1\rho}\right)\\
       & \quad \times \left(\slashed{q}_3+m_{B_c}\right) \left(g_{B_cD^*\Lambda}\gamma^\beta-i\frac{\kappa_{B_cD^*\Lambda}}{m_{B_c}+m_\Lambda}\sigma^{\beta\sigma}q_{2\sigma}\right) v(p_1)\\
       & \quad \times \frac{-g^\nu_\lambda+q_1^\nu q_{1\lambda}/m_{D^*}^2}{q_1^2-m_{D^*}^2} \frac{-g^\alpha_\beta+q_2^\alpha q_{2\beta}/m_{D^*}^2}{q_2^2-m_{D^*}^2} \frac{1}{q_3^2-m_{B_c}^2} \mathcal{F}^2(q_3^2).
   \end{split}
 \end{equation}
Here,  as form factor, $\mathcal{F}(q_3^2)$ is introduced to describe the off-shell effect of the exchanged charmed baryon in the rescattering process $D^{(*)}\bar{D}^{(*)}\to \Lambda\bar{\Lambda}$. In addition, introducing this form factor can avoid the divergence of the loop integral. In this work, we adopt the dipole form factor with expression
 \begin{equation}\label{eqalpha}
   \mathcal{F}(q_3^2) = \left( \frac{m_E^2-\Lambda^2}{q_3^2-\Lambda^2} \right)^2, \quad \Lambda = m_E+\alpha \Lambda_{QCD},
 \end{equation}
where $m_E$ and $q_3$ are the mass and four-momentum of the exchanged baryon, respectively. $\Lambda_{QCD} = 220$ MeV, and $\alpha$ is a phenomenological dimensionless parameter.

The total amplitude of the $Y(4230)\to\Lambda\bar{\Lambda}$ process reads as
 \begin{equation}
   \mathcal{M}\left(Y(4230)\to\Lambda\bar{\Lambda}\right) = 2\sum_{i=a,b,c,d}\left( \mathcal{M}^i_{\Xi_c}+\mathcal{M}^i_{\Xi_c'}\right) +\sum_{i=a,b,c,d}\mathcal{M}^i_{\Lambda_c},
 \end{equation}
where the factor of 2 in front of $\mathcal{M}^i_{\Xi_c}$ and $\mathcal{M}^i_{\Xi_c^\prime}$ comes from the summation over the isospin doublet $(D^{0(*)},D^{+(*)})$.

\begin{figure}[htbp]
  \centering
  \includegraphics[width=8cm,keepaspectratio]{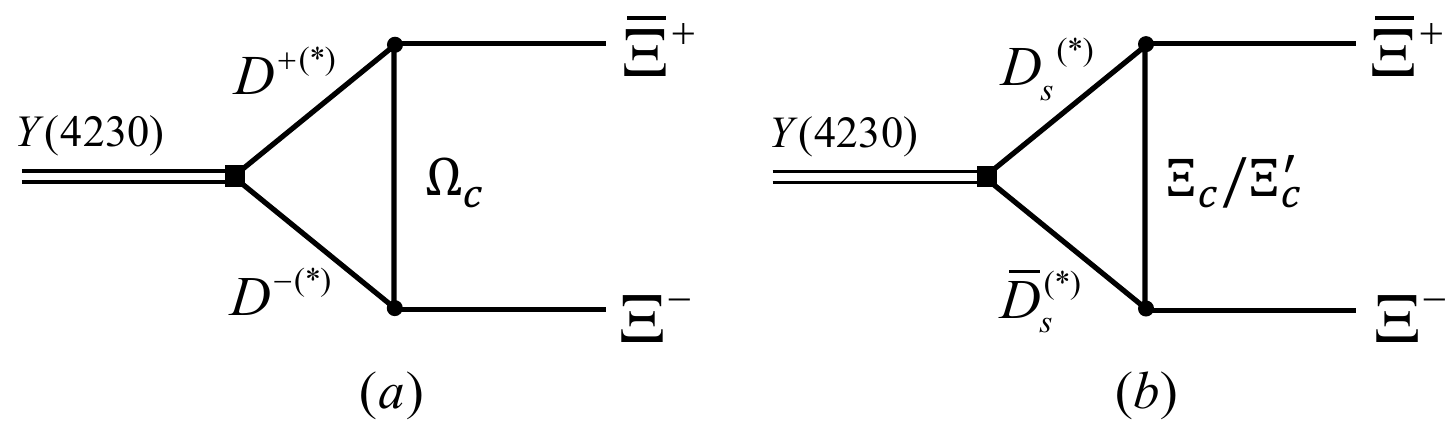}
  \caption{The schematic diagrams of the $Y(4230)\to\Xi^- \bar{\Xi}^+$ process induced by the hadronic loop mechanism.}\label{fig:Xi}
\end{figure}

In the case of the $Y(4230)\to\Xi^-\bar{\Xi}^+$ process, the intermediate charmed meson pairs can only be $D^{+(*)}D^{-(*)}$ and $D_s^{(*)}\bar{D}_s^{(*)}$. For the $D^{+(*)}D^{-(*)}$ intermediate state, the exchanged charmed baryon is $\Omega_c$. For the $D_s^{(*)}\bar{D}_s^{(*)}$ intermediate state, the exchanged charmed baryon can be either $\Xi_c$ or $\Xi_c^\prime$. The allowed Feynman diagrams are showed in Fig.~\ref{fig:Xi}.

To evaluate the diagrams given by Fig.~\ref{fig:Xi}, we need the effective Lagrangian of $B_cD^{(*)}_{(s)}\Xi^-$ couplings besides the $\mathcal{L}_{\psi\mathcal{D}^{(*)}\mathcal{D}^{(*)}}$ given in Eq.~\eqref{eq:lagrangianpsi}, which has the same form as the Lagrangian of $B_c D^{(*)}\Lambda$ interaction given in Eq.~\eqref{eq:lagrangianLa},
\begin{equation}\label{eq:lagrangianN}
   \begin{split}
     \mathcal{L}_{B_cD^{(*)}_{(s)}\Xi} &= ig_{B_cD_{(s)}\Xi}\bar{B}_c\gamma^5D_{(s)}\Xi \\
       & \quad +\bar{B}_c\left( g_{B_cD^*_{(s)}\Xi}\gamma^\mu + \frac{\kappa_{B_cD^*_{(s)}\Xi}}{m_{B_c}+m_\Xi}\sigma^{\mu\nu}\partial_\nu \right) D^*_{(s)\mu} \Xi +h.c..
   \end{split}
 \end{equation}
Similar to the case of the $Y(4230)\to\Lambda\bar{\Lambda}$ process, there are still four combinations of intermediate charmed meson pair as seen in Fig.~\ref{fig:Xi}. They have the same forms as those in Eqs.~\eqref{ampbeg}-\eqref{amped}. And the total amplitude of the  $Y(4230)\to\Xi^-\bar{\Xi}^+$ process reads as
 \begin{equation}
   \mathcal{M}\left(Y(4230)\to\Xi^-\bar{\Xi}^+\right) = \sum_{i=a,b,c,d}\left( \mathcal{M}^i_{\Omega_c}+\mathcal{M}^i_{\Xi_c}+\mathcal{M}^i_{\Xi_c^\prime}\right).
 \end{equation}

With these amplitudes, the branching ratio of $Y(4230)\to\Lambda\bar{\Lambda}/\Xi^-\bar{\Xi}^+$ can be calculated by
 \begin{equation}
   \mathcal{B} = \frac{1}{3}\frac{1}{8\pi} \frac{|\boldsymbol{p}_{\Lambda/\Xi^-}^{cm}|}{m_{Y}^2\Gamma_{Y}}\sum_{pol.}\left|\mathcal{M}\left(Y(4230)\to\Lambda\bar{\Lambda}/\Xi^-\bar{\Xi}^+\right)\right|^2,
 \end{equation}
where the factor of $1/3$ is due to spin average over an initial charmonium state.

\subsection{\label{results}The predicted branching ratios of $Y(4230) \to \Lambda \bar{\Lambda},\,\Xi^-\bar{\Xi}^+$}

To evaluate the branching ratios of the $Y(4230)\to\Lambda\bar{\Lambda}$ and $\Xi^-\bar{\Xi}^+$ processes, we need to fix the coupling constants in the effective Lagrangians. {Since we are discussing the process $Y\to\Lambda\bar{\Lambda}/\Xi^-\bar{\Xi}^+$ by assuming the $Y(4230)$ as a charmonium state, we may fix the coupling constants in Eq.~\eqref{eq:lagrangianpsi} by partial widths of the $Y(4230)$ given in Ref.~\cite{Wang:2019mhs}, where the author treat the $Y(4230)$ as a higher charmonium.} The partial widths and corresponding coupling constants are collected in Table~\ref{tab:psidd}.
 \begin{table*}
 \begin{ruledtabular}
   \centering
   \caption{Partial widths of the open-charm decay of the $Y(4230)$~\cite{Wang:2019mhs} and the obtained coupling constants. }\label{tab:psidd}
   \begin{tabular}{cccccc}
    Channel & $DD$ & $DD^*$ & $D^*D^*$ & $D_sD_s$ & $D_sD_s^*$  \\
     \hline
    Partial width (MeV) & 3.29 & 0.723 & 21.8 & $0.144$ & $0.0486$  \\
    Coupling constant & 0.765 & 0.054 $\mathrm{GeV}^{-1}$ & 1.320 & 0.233 & 0.027 $\mathrm{GeV}^{-1}$ \\
   \end{tabular}
   \end{ruledtabular}
 \end{table*}
The coupling constants of $B_cD^{(*)}\Lambda$ and $B_cD^{(*)}\Xi$ interactions are obtained from the couplings of $\Lambda_cDN$ and $\Sigma_cDN$ by applying SU(3) symmetry. Here, we use the $\Lambda_cDN$ and $\Sigma_cDN$ coupling constants obtained from the results of QCD light-cone sum rules as \cite{Khodjamirian:2011jp}
\begin{equation*}
    \begin{aligned}
        g_{\Lambda_cDN} &= 13.8, \\
        g_{\Lambda_cD^*N} &= -7.9, \\
        \kappa_{\Lambda_cD^*N} &= 4.7.
    \end{aligned}
    \qquad
    \begin{aligned}
        g_{\Sigma_cDN} &= 1.3, \\
        g_{\Sigma_cD^*N} &= 1.0, \\
        \kappa_{\Sigma_cD^*N} &= 2.1.
    \end{aligned}
\end{equation*}
And then, with SU(3) symmetry, the various couplings needed can be obtained by the following relations
\begin{align*}
  \sqrt{\frac{3}{2}}g_{\Lambda_cD_s^{(*)}\Lambda}=\sqrt{6}g_{\Xi_cD^{(*)}\Lambda}=g_{\Xi_cD_s^{(*)}\Xi}=g_{\Lambda_cD^{(*)}N},\\
  \sqrt{\frac{2}{3}}g_{\Xi_c'D^{(*)}\Lambda}=\frac{1}{\sqrt{2}}g_{\Omega_cD^{(*)}\Xi}=g_{\Xi_c^{\prime}D_s^{(*)}\Xi}=g_{\Sigma_cD^{(*)}N},
\end{align*}
where we omit possible minus signs in various coupling constants because they always appear twice in the amplitudes.

\begin{figure}
  \centering
  \includegraphics[width=80mm,keepaspectratio]{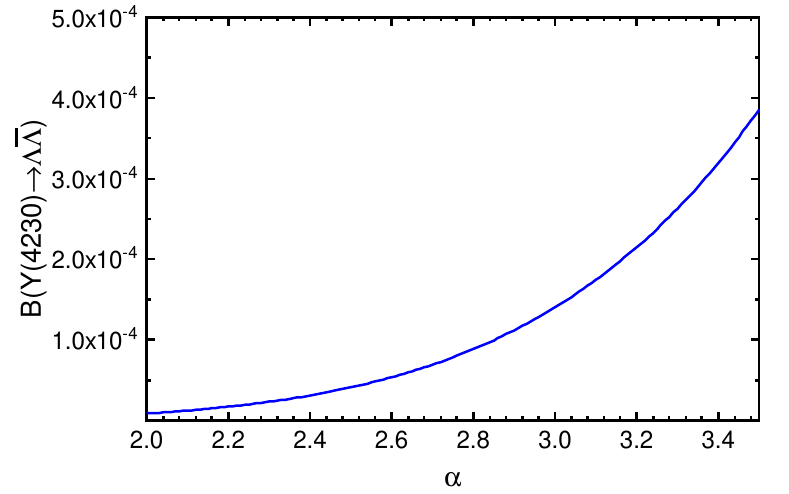}
  \caption{The $\alpha$ dependence of the branching ratio of $Y(4230)\to\Lambda\bar{\Lambda}$.}\label{fig:branching_La}
\end{figure}

\begin{figure}
  \centering
  \includegraphics[width=80mm,keepaspectratio]{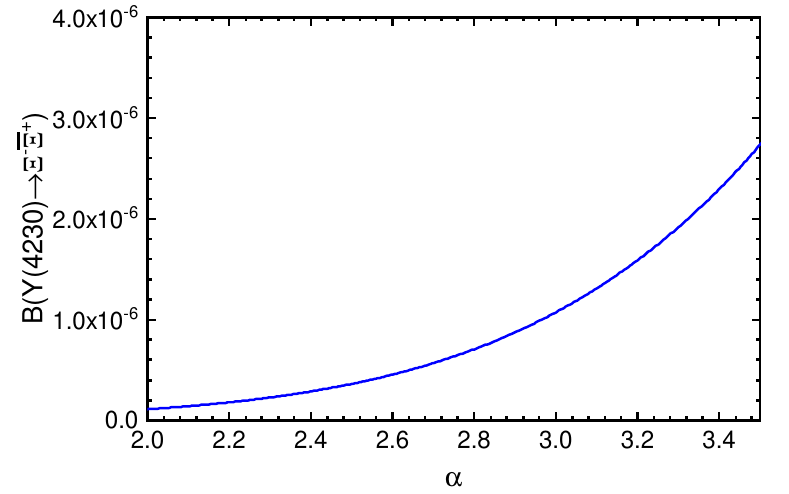}
  \caption{The $\alpha$ dependence of the branching ratio of $Y(4230)\to\Xi^-\bar{\Xi}^+$.}\label{fig:branching_Xi}
\end{figure}

There remains a cut-off parameter $\alpha$, which should be of order 1 and dependent on the specific process~\cite{Cheng:2004ru}. In fact, the hadronic loop mechanism had been applied to the study of the hidden-charm decays of the $Y(4230)$ \cite{Chen:2014sra,Chen:2015bma}, where the corresponding branching ratios like the $Y(4230)\to \chi_{c0}\omega$ and $\psi(3686)\pi^+\pi^-$ can be reproduced well. Thus, we can fix the $\alpha$ range by borrowing the experience of former studies~\cite{Chen:2014sra,Chen:2015bma} and take the $\alpha$ in the range $2-3.5$. The predicted branching ratios of $Y(4230)\to \Lambda\bar{\Lambda}$ and $\Xi^-\bar{\Xi}^+$ decays are shown in Fig.~\ref{fig:branching_La} and Fig.~\ref{fig:branching_Xi}, respectively. The resulting branching ratios are found to be in the range:
\begin{equation}\label{eq:branching_ratio_range}
  \begin{split}
     \mathcal{B}(Y(4230)\to\Lambda\bar{\Lambda}) & = 8.8\times10^{-6}-3.8\times10^{-4}, \\
     \mathcal{B}(Y(4230)\to\Xi^-\bar{\Xi}^+) & = 8.3\times10^{-8}-2.8\times10^{-6}.
  \end{split}
\end{equation}
{Here, we should indicate that the predicted values for the branching fractions in Ref.~\eqref{eq:branching_ratio_range} span almost two orders of magnitude. However, we can still make 
a few constraints on our results from the consideration of SU(3) symmetry, since the baryons $\Lambda$ and $\Xi$ belong to the same octet if considering SU(3) symmetry. Thus,   
assuming the $Y(4230)$ as a charmonium state, naive SU(3) symmetry anticipate the following relation
\begin{equation}
    \mathcal{B}(Y(4230)\to\Lambda\bar{\Lambda}):\mathcal{B}(Y(4230)\to\Xi^-\bar{\Xi}^+)\approx 1 .
\end{equation}
After comparing this relation to our calculated branching ratios, we suggest that our results favor the lower limit of $\mathcal{B}(Y(4230)\to\Lambda\bar{\Lambda})$ and the upper limit of $\mathcal{B}(Y(4230)\to\Xi^-\bar{\Xi}^+)$ in Eq.~\eqref{eq:branching_ratio_range}, both of which are of order $10^{-6}$.}
We hope that our experimental colleague may notice it and carry out the search for these decay modes of the $Y(4230)$. In the following, we illustrate the possibility of hunting for the $Y(4230)$ combining with the present experimental data.


\section{\label{discussion_data}A discussion of possibility of finding out $Y(4230)\to \Lambda \bar{\Lambda}$ and $\Xi^-\bar{\Xi}^+$ by combined with the BESIII data}
\subsection{\label{data}The BESIII's data of $Y(4230)\to \Lambda \bar{\Lambda}$ and $Y(4230)\Xi^-\bar{\Xi}^+$ }

\begin{figure}[htbp]
  \centering
  \includegraphics[width=8cm,keepaspectratio]{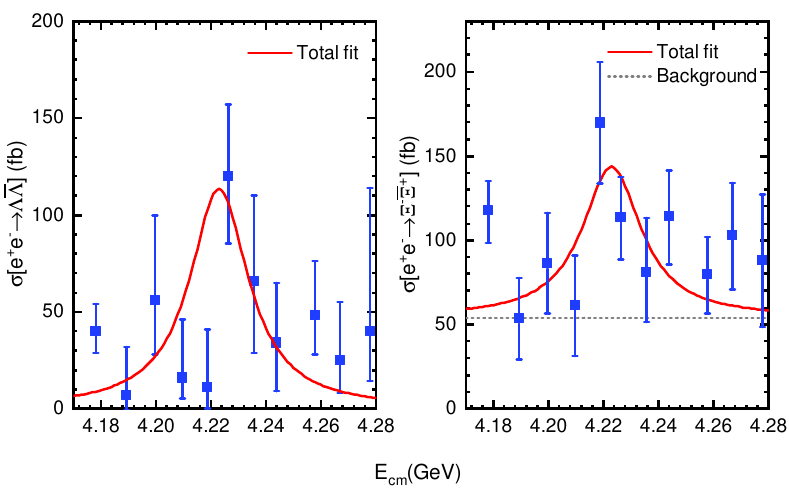}
  \caption{The possible signals of the $Y(4230)$ existing in the $e^+e^-\to\Lambda\bar{\Lambda}$ \cite{BESIII:2021ccp} and $e^+e^-\to\Xi^-\bar{\Xi}^+$ \cite{BESIII:2019cuv} channel. Here, for fitting the $e^+e^-\to \Xi^-\bar{\Xi}^+$ data, a constant background is used. }\label{fig:data}
\end{figure}

Recently, BESIII measured the $e^+e^-\to\Xi^-\bar{\Xi}^+$~\cite{BESIII:2019cuv} and $e^+e^-\to \Lambda\bar\Lambda$ \cite{BESIII:2021ccp} channels at different energy points, where the signals of the $Y(4230)$ are not significant. If carefully checking the data, event accumulations around $\sqrt{s}\approx4.22$ GeV can be found in these two channels (see Fig.~\ref{fig:data}), which should be caused by the potential $Y(4230)$ resonance.

Assuming the existence of the $Y(4230)$ in the $e^+e^-\to \Lambda\bar\Lambda$ \cite{BESIII:2021ccp} and the $e^+e^-\to \Xi^-\bar{\Xi}^+$ \cite{BESIII:2019cuv} processes, we try to do a simple fit to the experimental data by using the phase space corrected Breit-Wigner distribution:
\begin{equation}
  \mathcal{M}_R(Y) = \frac{\sqrt{12\pi\Gamma^{e^+e^-}_{Y} \mathcal{B}\left(Y\to\Lambda\bar{\Lambda}/\Xi^-\bar{\Xi}^+\right)\Gamma_{Y}}}{s-m^2_{Y}+im_{Y}\Gamma_{Y}} \sqrt{\frac{\Phi(s)}{\Phi(m_{\psi}^2)}},
\end{equation}
where $\Phi(s)$ denotes the two-body phase space. The mass and width of the $Y(4230)$ are taken from the calculated values in Ref.~\cite{Wang:2019mhs}, \textit{i.e.}, $m_{Y(4230)}=4220$ MeV, $\Gamma_{Y(4230)}=26$ MeV.
{The width we used is close to that measured in the $e^+e^-\to\chi_{c0}\omega$ channel, $\Gamma_Y^{\chi_{c0}\omega}=28.2$~\cite{BESIII:2019gjc}, which provide a clean environment to measure the resonance parameter because the threshold $M_{\chi_{c0}\omega}=4.196$ GeV suppresses possible interference from lower resonances like the $\psi(4040)$ and $\psi(4160)$.}
The results of these simple fits are shown in Fig.~\ref{fig:data}, from which we obtain the product of dilepton width and branching ratios as
\begin{equation} \label{eq:branchingratio_th}
  \begin{split}
    \Gamma_Y^{e^+e^-}\mathcal{B}(Y(4230)\to\Lambda\bar{\Lambda})& = (3.58\pm0.89)\times 10^{-6} \ \mathrm{keV}, \\
    \Gamma_Y^{e^+e^-}\mathcal{B}(Y(4230)\to\Xi^-\bar{\Xi}^+)& = (2.83\pm0.64) \times 10^{-6} \ \mathrm{keV}.
  \end{split}
\end{equation}
{In general, the dilepton width of S-wave charmonium is of order of 1 keV~\cite{ParticleDataGroup:2020ssz}. The dilepton width of $\psi(4S)$ was calculated to be 0.63 keV in Ref.~\cite{Dong:1994zj}, and 0.63 keV in Ref.~\cite{Li:2009zu}, the mass of $\psi(4S)$ is around 4.2 GeV in both of those paper. In Ref.~\cite{Wang:2019mhs}, the charmonium assignment of $Y(4230)$ was considered in detail, and the dilepton width was found to be 0.29 keV. Here, we take $\Gamma_Y^{e^+e^-}=0.29$ keV, then the extracted branching ratios are:}
\begin{equation} \label{eq:branchingratio_exp}
  \begin{split}
    \mathcal{B}(Y(4230)\to\Lambda\bar{\Lambda})& = (12.35\pm3.06)\times 10^{-6}, \\
    \mathcal{B}(Y(4230)\to\Xi^-\bar{\Xi}^+)& = (9.77\pm2.19) \times 10^{-6}.
  \end{split}
\end{equation}

We may take these extracted values to make further discussion associated with our calculated result in Sec. \ref{results}. 
The lower limit of our calculated results of $\mathcal{B}(Y(4230)\to\Lambda\bar{\Lambda})$ shown in Eq.~\eqref{eq:branching_ratio_range} 
can reach up to the extracted value from BESIII's data, while upper limit of the  calculated  $\mathcal{B}(Y(4230)\to\Xi^-\bar{\Xi}^+)$ can meet the extracted data mentioned above. {We also notice that indeed these extracted values of $\mathcal{B}(Y(4230)\to\Lambda\bar{\Lambda})$ and $\mathcal{B}(Y(4230)\to\Xi^-\bar{\Xi}^+)$ by the BESIII's data
satisfy the requirement of the SU(3) symmetry.}
This fact enforce our ambition of suggesting experimentalist to search for the $Y(4230)$ signal via the $e^+e^-\to \Lambda\bar{\Lambda}$ and $e^+e^-\to \Xi^-\bar{\Xi}^+$, which may become a potential issue of BESIII.

\subsection{\label{formfactors} The EMFFs of $\Lambda$ and $\Xi^-$ at $\sqrt{s}=m_{Y(4230)}$  }

As mentioned in the introduction, the time-like EMFFs at specific resonance point reflect the complete amplitude of the resonance decay. However, the decay branching ratio only utilize the module of the amplitude. So in this section, we explore the EMFFs of $\Lambda$ and $\Xi^-$ at $\sqrt{s}=m_{Y(4230)}$, which is an observables different from the total cross section data and can be measured in future experiment.

The photon interacting with baryon and anti-baryon is absorbed into the vertex $\gamma B\bar{B}$, which can be described by the so-called Sachs EMFFs $G_E$ and $G_M$~\cite{Ernst:1960zza}. In terms of $G_E$ and $G_M$, the $\gamma B\bar{B}$ vertex can be expressed as
\begin{equation}\label{eq:ff}
  -ie\Gamma^\mu=-ie\bar{u}(p_2)\left( G_M\gamma^\mu-\frac{2m_\Lambda}{Q^2}\left( G_M-G_E \right)Q^\mu \right)v(p_1),
\end{equation}
where $Q=p_2-p_1$, $G_E$ and $G_M$ are the electric and magnetic form factors,  respectively. The form factors $G_E$ and $G_M$ are analytical functions of the four-momentum transfer squared $q^2$, analytic in the $q^2$-complex plane with a branch cut along the positive real axis because of possible hadronic channels coupled with the virtual photon and the $B\bar{B}$ state. In the space-like region $q^2<0$, both $G_E$ and $G_M$ are real values. In the time-like region $q^2=s>(2m_B)^2$, the form factors are complex numbers because of the hadronic channels coupled with both of photon and $B\bar{B}$ state
\footnote{Actually, we are discussing the region $q^2+i\epsilon$ since the physical scattering amplitudes are evaluated above the branch cut.}.

{The space-like EMFFs of stable baryons such as proton can be measured through the scattering process $e^-B\to e^-B$. For unstable baryons such as $\Lambda$ and $\Xi^-$ hyperon, this scattering process cannot be accessible at present.} However, the time-like $G_E$ and $G_M$ of hyperons can be extracted from the  $e^+e^-$ annihilation processes $e^+e^-\to B\bar{B}$. Especially, the relative phase $\Delta\Phi=\arg(G_E/G_M)$ is related to the polarization of the produced baryon or antibaryon. Directly measuring the polarization of a baryon is not realistic. An alternate approach to obtain the polarization of a hyperon is through the angular distribution of the subsequent decay of a hyperon like $\Xi\to\Lambda\pi$ and $\Lambda\to p\pi$. Recently, several measurements of the complete EMFFs of $\Lambda$ and $\Xi^-$ have been done by BESIII. The module and relative phase of the EMFFs of $\Lambda$ were given at $\sqrt{s}=m_{J/\psi}$ \cite{BESIII:2018cnd} and $\sqrt{s}=2.396$ GeV \cite{BESIII:2019nep}. The EMFFs of $\Xi^-$ were measured at $\sqrt{s}=m_{J/\psi}$ \cite{BESIII:2021ypr}.

\begin{figure}[htbp]
  \centering
  \includegraphics[width=8cm,keepaspectratio]{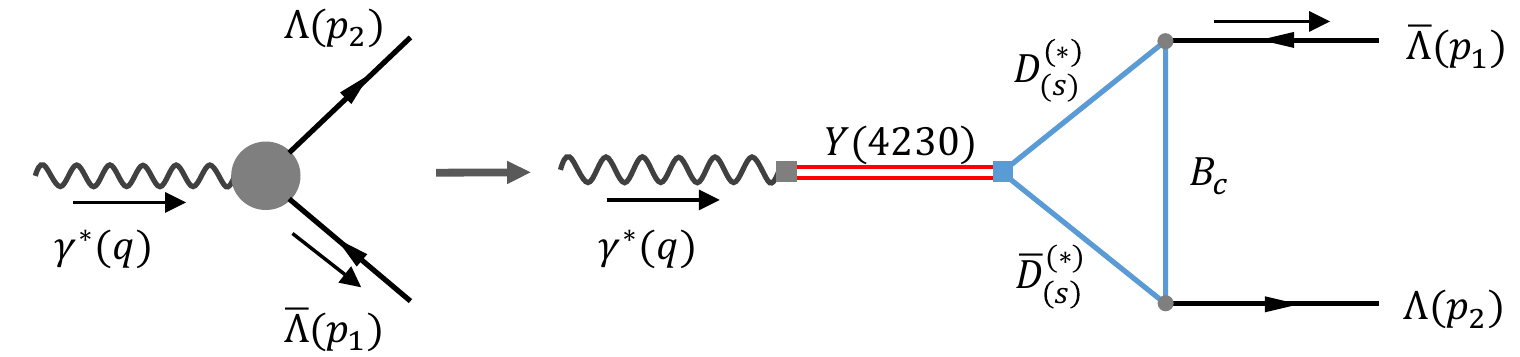}
  \caption{$\gamma\to \Lambda\bar{\Lambda}$ under the hadronic loop mechanism. Here, energy point is fixed to be $\sqrt{s}=m_{Y(4230)}$ and $B_c$ denotes these allowed charmed baryons.}\label{fig:ff}
\end{figure}

Assuming that the $Y(4230)$ exist in $e^+e^-\Lambda\bar{\Lambda}$ and $e^+e^-\Xi^-\bar{\Xi}^+$ process, then at $\sqrt{s}=m_{Y(4230)}$, the process we considered is actually $e^+e^-\to\gamma^*\to Y(4230)\to\Lambda\bar{\Lambda}/\Xi^-\bar{\Xi}^+$. The $\gamma^*\Lambda\bar{\Lambda}$ and $\gamma^*\Xi^-\bar{\Xi}^+$ vertices are dominated by the hadronic intermediate states, where the triangle loops composed of charmed mesons may have connection with both the $Y(4230)$ and the final state $\Lambda\bar{\Lambda}/\Xi^-\bar{\Xi}^+$ (see Fig.~\ref{fig:ff}). Then EMFFs of $\Lambda\bar{\Lambda}$ and $\Xi^-\bar{\Xi}^+$ can reflect the information of these hadronic loops.
The loop integral has both real and imaginary parts corresponding to to the complex EMFFs exactly.

The amplitudes of these Feynman diagrams in Fig.~\ref{fig:ff} are almost the same as those in the previous section, the only difference is that we need a coupling $g_Y$ to describe the transition amplitude of a photon to $Y(4230)$, which is related to the dilepton width of $Y(4230)$ by the relation:
\begin{equation}
  g_Y^2 = \frac{12\pi m_Y^3 \Gamma^{e^+e^-}_Y}{e^2}.
\end{equation}
Writing out the amplitude of $Y(4230)\to\Lambda\bar{\Lambda}/\Xi^-\bar{\Xi}^+$ in the form $\mathcal{M}=\epsilon_\psi^\mu \mathcal{M}_\mu^Y$, then the $\gamma^*B\bar{B}$ vertex is further expressed as
\begin{equation}\label{ampff}
  \mathcal{M}^\mu_{\gamma^*B\bar{B}}=g_Y \frac{-g^{\mu\kappa}+q^\mu q^\kappa /m_Y^2}{q^2-m_Y^2+im_Y\Gamma_Y}\mathcal{M}_\kappa^Y.
\end{equation}
After evaluating the loop integral in the amplitude, this vertex can be reduced to the form of Eq.~\eqref{eq:ff}, from which we can get the result of the form factors $G_E$ and $G_M$.

\begin{figure}[htbp]
  \centering
  \includegraphics[width=8cm,keepaspectratio]{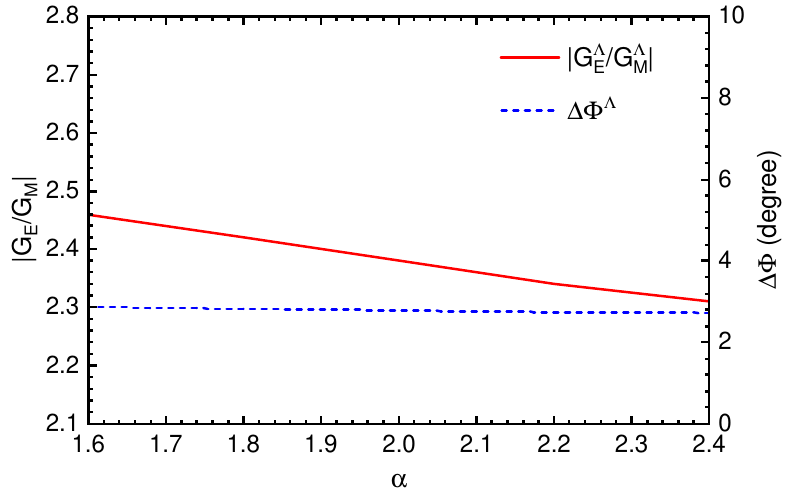}
  \caption{The $\alpha$ dependence of the $|G_E/G_M|$ and $\Delta \Phi$ of $\Lambda$ at $\sqrt{s}=m_{Y(4230)}$.}\label{fig:ffla}
\end{figure}

\begin{figure}[htbp]
  \centering
  \includegraphics[width=8cm,keepaspectratio]{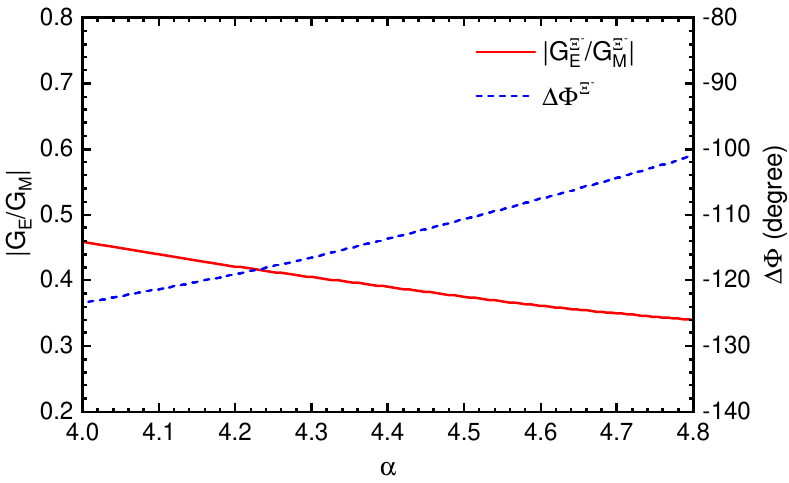}
  \caption{ The $\alpha$ dependence of the $|G_E/G_M|$ and $\Delta \Phi$ of $\Xi^-$ at $\sqrt{s}=m_{Y(4230)}$.}\label{fig:ffxi}
\end{figure}

In Fig.~\ref{fig:ffla} and Fig.~\ref{fig:ffxi}, we show the obtained EMFFs of $\Lambda$ and $\Xi^-$, respectively. The cut-off parameter $\alpha$ are taken to be around 2 and $4.4$ for $\Lambda\bar{\Lambda}$ and $\Xi^-\bar{\Xi}^+$ channels so as to produce the extracted branching ratios in Eq.~\eqref{eq:branchingratio_exp}.
The behavior of the $|G_E/G_M|$ and $\Delta \Phi$ for $\Lambda$ and $\Xi^-$ are quite different. Here, the $|G_E/G_M|$ of $\Lambda$ is about 2.4 while that of $\Xi^-$ is around 0.4. In addition, the $\Delta \Phi$ of $\Lambda$ has small value, while  the $\Delta \Phi$ of $\Xi^-$ has a much larger absolute value, and has a minus sign. For these two processes, the $|G_E/G_M|$ and $\Delta \Phi$ have weak dependence on $\alpha$, where the predicted ratios of EMFFs are stable.
The results of EMFFs of $\Lambda$ and $\Xi^-$ when taking typical values $\alpha_\Lambda=2.10$ and $\alpha_{\Xi^-}=4.34$ are collected in Table~\ref{tab:emffs}.
We suggest future experiment like BESIII and BelleII to perform the measurement of these EMFFs, by which the charmonia assignment to the $Y(4230)$ in addtion with the hadronic loop mechanism itself can be further tested.


 \begin{table}[htbp]
 \begin{ruledtabular}
   \centering
   \caption{ The predicted ratios of EMFFs of $\Lambda$ and $\Xi^-$ at $\sqrt{s}=m_{Y(4230)}$.}\label{tab:emffs}
   \begin{tabular}{ccccc}
    Form factor & $|G_E^\Lambda/G_M^\Lambda|$ & $\Delta \Phi^\Lambda$ & $|G_E^{\Xi^-}/G_M^{\Xi^-}|$ & $\Delta \Phi^{\Xi^-}$ \\
    \hline
    Value & 2.36 & $2.74^\circ$ & 0.40 & $-115.60^\circ$
   \end{tabular}
   \end{ruledtabular}
 \end{table}

\section{Summery}

With the accumulation of experimental data, charmoniumlike state $Y(4230)$
was observed by BESIII in $e^+e^-\to J/\psi\pi^+\pi^-$ \cite{BESIII:2016bnd}, $e^+e^-\to h_c\pi^+\pi^-$ \cite{BESIII:2016adj}, $e^+e^-\to \chi_{c0}\omega$ \cite{BESIII:2014rja}, and $e^+e^-\to \psi(3686)\pi^+\pi^-$ \cite{BESIII:2017tqk}, which is consistent with the predicted vector charmonium state around 4.2 GeV \cite{He:2014xna}. As indicated by the recent work, the $Y(4230)$ can be assigned to be the mixture of the $\psi(4S)$ and $\psi(3D)$ charmonium states \cite{Wang:2019mhs}. What is more important is that the $Y(4230)$ may play the role of scaling point when constructing the $J/\psi$ family.

In fact, the hidden-charm decay is only one mode of these allowed decays of the $Y(4230)$ under the vector charmonium assignment to the $Y(4230)$. Later, the BESIII observation of an enhancement around 4.2 GeV in the $e^+e^-\to \pi^+ D^0 D^{*-}$ enforces the above conclusion of decoding the $Y(4260)$ as charmonium
\cite{BESIII:2018iea}.

If checking the decay behaviors of these established charmonia like $J/\psi$, $\psi(3686)$ and $\psi(3770)$, we find their abundant decay modes \cite{ParticleDataGroup:2020ssz}. Besides hidden-charm and open-charm decays, there exist the decays into light hadrons for charmonium.

Inspired by the above situation, we propose to study the $Y(4230)$ decays into $\Lambda\bar{\Lambda}$ and $\Xi^-\bar{\Xi}^+$, which is also due to the experimental potential of analyzing these decays \cite{BESIII:2021ccp,BESIII:2019cuv}.
Assuming the $Y(4230)$ as a charmonium state, we calculated the branching ratos of these two channels via hadronic loop mechanism. The branching ratios of $\Lambda\bar{\Lambda}$ and $\Xi^-\bar{\Xi}^+$ channels are found to be of order $10^{-6}$.
We notice the present experimental data of 
$e^+e^-\to\Lambda\bar{\Lambda}$ and $e^+e^-\to \Xi^-\bar{\Xi}^+$ given by BESIII. We discuss the possibility of searching for the $Y(4230)$ signal via these two processes. We strongly suggest our experimental colleague to perform the precise measurement of $e^+e^-\to \Lambda\bar{\Lambda}$ and $\Xi^-\bar{\Xi}^+$ in future.

Based on the assumption that $Y(4230)$ exist in the present data, we further predict the EMFFs of $\Lambda$ and $\Xi^-$ at $\sqrt{s}=m_{Y(4230)}$. The $|G_E^\Lambda /G_M^\Lambda|$ and $\Delta\Phi^\Lambda$ are predicted to be 2.36 and $2.74^\circ$, respectively. And, the $|G_E^{\Xi^-}/G_M^{\Xi^-}|$ and $\Delta\Phi^{\Xi^-}$ are predicted to be 0.40 and $-115.60^\circ$, respectively. These predictions on EMFFs can be accessible at future experiment like BESIII and BelleII.

Very recently, BESIII reported the measurement of EMFFs at $\sqrt{s}=3.773$ GeV~\cite{BESIII:2021cvv} which is in the vicinity $\psi(3770)$ peak. The measurement of EMFFs at $Y(4230)$ point will be an interesting topic in the near future. We hope that future experiments like BESIII and BelleII can seriously focus on this issue with higher precision.

\begin{acknowledgments}
This work is supported by the China National Funds for Distinguished Young Scientists under Grant No. 11825503, National Key Research and Development Program of China under Contract No. 2020YFA0406400, the 111 Project under Grant No. B20063, the National Natural Science Foundation of China under Grant No. 12047501, and by the Fundamental Research Funds for the Central Universities.
\end{acknowledgments}



\begin{thebibliography}{90}

\bibitem{Aubert:1974js}
  J.~J.~Aubert \textit{et al.} (E598 Collaboration),
  Experimental Observation of a Heavy Particle $J$,
  \href{http://dx.doi.org/10.1103/PhysRevLett.33.1404}{Phys.\ Rev.\ Lett.\ \textbf{33}, 1404 (1974)}.

\bibitem{Augustin:1974xw}
  J.~E.~Augustin \textit{et al.} (SLAC-SP-017 Collaboration),
  Discovery of a Narrow Resonance in $e^+ e^-$ Annihilation,
  \href{http://dx.doi.org/10.1103/PhysRevLett.33.1406}{Phys.\ Rev.\ Lett.\ \textbf{33}, 1406 (1974)}.

\bibitem{Liu:2019zoy}
  Y.~R.~Liu, H.~X.~Chen, W.~Chen, X.~Liu, and S.~L.~Zhu,
  Pentaquark and Tetraquark states,
  \href{http://dx.doi.org/10.1016/j.ppnp.2019.04.003}{Prog.\ Part.\ Nucl.\ Phys.\ \textbf{107}, 237 (2019)}.

\bibitem{Brambilla:2019esw}
  N.~Brambilla, S.~Eidelman, C.~Hanhart, A.~Nefediev, C.~P.~Shen, C.~E.~Thomas, A.~Vairo, and C.~Z.~Yuan,
  The $XYZ$ states: experimental and theoretical status and perspectives,
  \href{http://dx.doi.org/10.1016/j.physrep.2020.05.001}{Phys.\ Rept.\ \textbf{873}, 1 (2020)}.

\bibitem{He:2014xna}
  L.~P.~He, D.~Y.~Chen, X.~Liu, and T.~Matsuki,
  Prediction of a missing higher charmonium around 4.26 GeV in $J/\psi$ family,
  \href{http://dx.doi.org/10.1140/epjc/s10052-014-3208-5}{Eur.\ Phys.\ J.\ C {\bf 74}, 3208 (2014)}.

\bibitem{Li:2009zu}
  B.~Q.~Li and K.~T.~Chao,
  Higher Charmonia and $X$, $Y$, $Z$ states with Screened Potential,
  \href{http://dx.doi.org/10.1103/PhysRevD.79.094004}{Phys.\ Rev.\ D {\bf 79}, 094004 (2009)}.

\bibitem{Yuan:2013uta}
  C.~Z.~Yuan,
  Evidence for resonant structures in $e^{+}e^{-} \to \pi^{+}\pi^{-}h_c$,
  \href{http://dx.doi.org/10.1088/1674-1137/38/4/043001}{Chin.\ Phys.\ C \textbf{38}, 043001 (2014)}.

\bibitem{BESIII:2013ouc}
  M.~Ablikim \textit{et al.} (BESIII Collaboration),
  Observation of a Charged Charmoniumlike Structure $Z_c$(4020) and Search for the $Z_c$(3900) in $e^+e^- \to \pi^+\pi^-h_c$,
  \href{http://dx.doi.org/10.1103/PhysRevLett.111.242001}{Phys.\ Rev.\ Lett.\ \textbf{111}, 242001 (2013)}.

\bibitem{Belle:2013yex}
  Z.~Q.~Liu \textit{et al.} (Belle Collaboration),
  Study of $e^+e^- \to \pi^+ \pi^- J/\psi$ and Observation of a Charged Charmoniumlike State at Belle,
  \href{http://dx.doi.org/doi:10.1103/PhysRevLett.110.252002}{Phys.\ Rev.\ Lett.\ \textbf{110}, 252002 (2013)}.

\bibitem{Ablikim:2014qwy}
  M.~Ablikim {\it et al.} (BESIII Collaboration),
  Study of $e^+e^-\to\omega\chi_{cJ}$ at center-of-mass energies from 4.21 to 4.42 GeV,
  \href{http://dx.doi.org/10.1103/PhysRevLett.114.092003}{Phys.\ Rev.\ Lett.\  {\bf 114}, 092003 (2015)}.

\bibitem{Chen:2014sra}
  D.~Y.~Chen, X.~Liu, and T.~Matsuki,
  Observation of $e^+e^-\to \chi_{c0}\omega$ and missing higher charmonium $\psi(4S)$,
  \href{http://dx.doi.org/10.1103/PhysRevD.91.094023}{Phys.\ Rev.\ D \textbf{91}, 094023 (2015)}.

\bibitem{Chen:2015bma}
  D.~Y.~Chen, X.~Liu, and T.~Matsuki,
  Search for missing $\psi(4S)$ in the $e^+e^-\to \pi^+\pi^-\psi(2S)$ process,
  \href{http://dx.doi.org/10.1103/PhysRevD.93.034028}{Phys.\ Rev.\ D \textbf{93}, 034028 (2016)}

\bibitem{Belle:2014wyt}
  X.~L.~Wang \textit{et al.} (Belle Collaboration),
  Measurement of $e^+e^- \to \pi^+\pi^-\psi(2S)$ via Initial State Radiation at Belle,
  \href{http://dx.doi.org/10.1103/PhysRevD.91.112007}{Phys.\ Rev.\ D \textbf{91}, 112007 (2015)}.

\bibitem{BESIII:2014rja}
  M.~Ablikim \textit{et al.} (BESIII Collaboration),
  Study of $e^+e^-\to\omega\chi_{cJ}$ at center-of-mass energies from 4.21 to 4.42 GeV,
  \href{http://dx.doi.org/10.1103/PhysRevLett.114.092003}{Phys.\ Rev.\ Lett.\ \textbf{114}, 092003 (2015)}.

\bibitem{BESIII:2017tqk}
  M.~Ablikim \textit{et al.} (BESIII Collaboration),
  Measurement of $e^{+}e^{-}\rightarrow \pi^{+}\pi^{-}\psi(3686)$ from 4.008 to 4.600\textasciitilde{}GeV and observation of a charged structure in the $\pi^{\pm}\psi(3686)$ mass spectrum,
  \href{http://dx.doi.org/10.1103/PhysRevD.96.032004}{Phys.\ Rev.\ D \textbf{96}, 032004 (2017); \textbf{99}, 019903(E) (2019)}.

\bibitem{BESIII:2018iea}
  M.~Ablikim \textit{et al.} (BESIII Collaboration),
  Evidence of a resonant structure in the $e^+e^-\to \pi^+D^0D^{*-}$ cross section between 4.05 and 4.60 GeV,
  \href{http://dx.doi.org/10.1103/PhysRevLett.122.102002}{Phys.\ Rev.\ Lett.\ \textbf{122}, 102002 (2019)}.

\bibitem{Wang:2019mhs}
  J.~Z.~Wang, D.~Y.~Chen, X.~Liu, and T.~Matsuki,
  Constructing $J/\psi$ family with updated data of charmoniumlike $Y$ states,
  \href{http://dx.doi.org/10.1103/PhysRevD.99.114003}{Phys.\ Rev.\ D \textbf{99}, 114003 (2019)}.

\bibitem{BESIII:2016bnd}
  M.~Ablikim \textit{et al.} (BESIII Collaboration),
  Precise measurement of the $e^+e^-\to \pi^+\pi^-J/\psi$ cross section at center-of-mass energies from 3.77 to 4.60 GeV,
  \href{http://dx.doi.org/10.1103/PhysRevLett.118.092001}{Phys.\ Rev.\ Lett.\ \textbf{118}, 092001 (2017)}.

\bibitem{BESIII:2016adj}
  M.~Ablikim \textit{et al.} (BESIII Collaboration),
  Evidence of Two Resonant Structures in $e^+ e^- \to \pi^+ \pi^- h_c$,
  \href{http://dx.doi.org/10.1103/PhysRevLett.118.092002}{Phys.\ Rev.\ Lett.\ \textbf{118}, 092002 (2017)}.

\bibitem{Cleven:2013mka}
  M.~Cleven, Q.~Wang, F.~K.~Guo, C.~Hanhart, U.~G.~Mei\ss{}ner and Q.~Zhao,
  $Y(4260)$ as the first $S$-wave open charm vector molecular state?,
  \href{http://dx.doi.org/10.1103/PhysRevD.90.074039}{Phys. Rev. D \textbf{90}, 074039 (2014)}

\bibitem{Ding:2008gr}
  G.~J.~Ding,
  Are $Y(4260)$ and $Z^+_2$ are $D_1D$ or $D_0D*$ Hadronic Molecules?,
  \href{http://dx.doi.org/10.1103/PhysRevD.79.014001}{Phys. Rev. D \textbf{79}, 014001 (2009)}
\bibitem{Zhu:2005hp}
S.~L.~Zhu,
  The Possible interpretations of $Y(4260)$,
  \href{http://dx.doi.org/10.1016/j.physletb.2005.08.068}{Phys.\ Lett.\ B\ \textbf{625}, 212 (2005)}

\bibitem{Close:2005iz}
F.~E.~Close and P.~R.~Page,
  Gluonic charmonium resonances at BaBar and BELLE?,
  \href{http://dx.doi.org/10.1016/j.physletb.2005.09.016}{Phys.\ Lett.\ B\ \textbf{628}, 215 (2005)}

\bibitem{Kou:2005gt}
E.~Kou and O.~Pene,
  Suppressed decay into open charm for the $Y(4260)$ being an hybrid,
  \href{http://dx.doi.org/10.1016/j.physletb.2005.09.013}{Phys.\ Lett.\ B\ \textbf{631}, 164 (2005)}

\bibitem{Maiani:2014aja}
L.~Maiani, F.~Piccinini, A.~D.~Polosa and V.~Riquer,
  The $Z(4430)$ and a New Paradigm for Spin Interactions in Tetraquarks,
  \href{http://dx.doi.org/10.1103/PhysRevD.89.114010}{Phys.\ Rev.\ D\ \textbf{89}, 114010 (2014)}

\bibitem{Dubynskiy:2008mq}
  S.~Dubynskiy and M.~B.~Voloshin,
  Hadro-Charmonium,
  \href{http://dx.doi.org/10.1016/j.physletb.2008.07.086}{Phys.\ Lett.\ B\ \textbf{666}, 344 (2008)}

\bibitem{Li:2013ssa}
  X.~Li and M.~B.~Voloshin,
  $Y$(4260) and $Y$(4360) as mixed hadrocharmonium,
  \href{http://dx.doi.org/10.1142/S0217732314500606}{Mod.\ Phys.\ Lett.\ A\ \textbf{29}, 1450060 (2014)}

\bibitem{ParticleDataGroup:2020ssz}
  P.~A.~Zyla \textit{et al.} (Particle Data Group),
  Review of Particle Physics,
  \href{http://dx.doi.org/10.1093/ptep/ptaa104}{PTEP\ \textbf(2020), 083C01 (2020)}


\bibitem{BaBar:2005hhc}
  B.~Aubert \textit{et al.} (BaBar Collaboration),
  Observation of a broad structure in the $\pi^+ \pi^- J/\psi$ mass spectrum around 4.26 GeV/c$^2$,
  \href{http://dx.doi.org/10.1103/PhysRevLett.95.142001}{Phys.\ Rev.\ Lett.\ \textbf{95}, 142001 (2005)}
  doi:10.1103/PhysRevLett.95.142001
  [arXiv:hep-ex/0506081 [hep-ex]].
  
\bibitem{BESIII:2019qvy}
  M.~Ablikim \textit{et al.} (BESIII Collaboration),
  Study of $e^+e^- \to \gamma \omega J/\psi$ and Observation of $X(3872) \to \omega J/\psi$,
  \href{http://dx.doi.org/10.1103/PhysRevLett.122.232002}{Phys.\ Rev.\ Lett.\ \textbf{122}, 232002 (2019)}

\bibitem{BESIII:2021ccp}
  M.~Ablikim \textit{et al.} (BESIII Collaboration),
  Measurement of the Cross Section for $e^{+}e^{-}\rightarrow\Lambda\bar\Lambda$ and Observation of the Decay $\psi(3770)\rightarrow\Lambda\bar\Lambda$,
  \href{http://arxiv.org/abs/2108.02410}{arXiv: 2108.02410}.

\bibitem{BESIII:2019cuv}
  M.~Ablikim \textit{et al.} (BESIII Collaboration),
  Measurement of the cross section for $e^{+}e^{-}\rightarrow\Xi^{-}\bar\Xi^{+}$ and observation of an excited $\Xi$ baryon,
  \href{http://dx.doi.org/10.1103/PhysRevLett.124.032002}{Phys.\ Rev.\ Lett.\ \textbf{124}, 032002 (2020)}.

\bibitem{BESIII:2018cnd}
  M.~Ablikim \textit{et al.} (BESIII Collaboration),
  Polarization and Entanglement in Baryon-Antibaryon Pair Production in Electron-Positron Annihilation,
  \href{http://dx.doi.org/10.1038/s41567-019-0494-8}{Nature Phys.\ \textbf{15}, 631 (2019)}.

\bibitem{BESIII:2019nep}
  M.~Ablikim \textit{et al.} (BESIII Collaboration),
  Complete Measurement of the $\Lambda$ Electromagnetic Form Factors,
  \href{http://dx.doi.org/10.1103/10.1103/PhysRevLett.123.122003}{Phys.\ Rev.\ Lett.\ \textbf{123}, 122003 (2019)}.

\bibitem{BESIII:2021ypr}
  M.~Ablikim \textit{et al.} (BESIII Collaboration),
  Weak phases and CP-symmetry tests in sequential decays of entangled double-strange baryons,
 \href{http://arxiv.org/abs/2105.11155}{arXiv: 2105.11155}.

\bibitem{BESIII:2020fqg}
  M.~Ablikim \textit{et al.} (BESIII Collaboration),
  $\Sigma^{+}$ and $\bar{\Sigma}^-$ polarization in the $J/\psi$ and $\psi(3686)$ decays,
  \href{http://dx.doi.org/10.1103/PhysRevLett.125.052004}{Phys.\ Rev.\ Lett. \textbf{125}, 052004 (2020)}.

\bibitem{Liu:2006dq}
  X.~Liu, X.~Q.~Zeng, and X.~Q.~Li,
  Study on contributions of hadronic loops to decays of $J/\psi \to$ vector + pseudoscalar mesons,
  \href{http://dx.doi.org/10.1103/PhysRevD.74.074003}{Phys.\ Rev.\ D \textbf{74}, 074003 (2006)}.

\bibitem{Liu:2009dr}
  X.~Liu, B.~Zhang, and X.~Q.~Li,
  The Puzzle of excessive non-$D\overline{D}$ component of the inclusive $\psi(3770)$ decay and the long-distant contribution,
  \href{http://dx.doi.org/10.1016/j.physletb.2009.04.047}{Phys.\ Lett.\ B \textbf{675}, 441 (2009)}.

\bibitem{Zhang:2009kr}
  Y.~J.~Zhang, G.~Li, and Q.~Zhao,
  Further understanding of the non-$D\overline{D}$ decays of $\ensuremath{\psi}(3770)$,
  \href{http://dx.doi.org/10.1103/PhysRevLett.102.172001}{Phys.\ Rev.\ Lett.\ \textbf{102}, 172001 (2009)}.

\bibitem{Meng:2007tk}
  C.~Meng and K.~T.~Chao,
  Scalar resonance contributions to the dipion transition rates of $\ensuremath{\Upsilon}(4S,5S)$ in the rescattering model,
  \href{http://dx.doi.org/10.1103/PhysRevD.77.074003}{Phys.\ Rev.\ D \textbf{77}, 074003 (2008)}.

\bibitem{Meng:2008dd}
  C.~Meng and K.~T.~Chao,
  Peak shifts due to ${B}^{(*)}\ensuremath{-}{\overline{B}}^{(*)}$ rescattering in $\ensuremath{\Upsilon}(5S)$ dipion transitions,
  \href{http://dx.doi.org/10.1103/PhysRevD.78.034022}{Phys.\ Rev.\ D \textbf{78}, 034022 (2008)}.

\bibitem{Meng:2008bq}
  C.~Meng and K.~T.~Chao,
  $\ensuremath{\Upsilon}(4S,5S)\ensuremath{\rightarrow}\ensuremath{\Upsilon}(1S)\ensuremath{\eta}$ transitions in the rescattering model and the new BABAR measurement,
  \href{http://dx.doi.org/10.1103/PhysRevD.78.074001}{Phys.\ Rev.\ D \textbf{78}, 074001 (2008)}.

\bibitem{Chen:2011qx}
  D.~Y.~Chen, J.~He, X.~Q.~Li, and X.~Liu,
  Dipion invariant mass distribution of the anomalous $\Upsilon(1S) \pi^{+} \pi^{-}$ and $\Upsilon(2S) \pi^{+} \pi^{-}$ production near the peak of $\Upsilon(10860)$,
  \href{http://dx.doi.org/10.1103/PhysRevD.84.074006}{Phys.\ Rev.\ D \textbf{84}, 074006 (2011)}.

\bibitem{Chen:2011zv}
  D.~Y.~Chen, X.~Liu, and S.~L.~Zhu,
  Charged bottomonium-like states $Z_b(10610)$ and $Z_b(10650)$ and the $\Upsilon(5S)\to \Upsilon(2S)\pi^+\pi^-$ decay,
  \href{http://dx.doi.org/10.1103/PhysRevD.84.074016}{Phys.\ Rev.\ D \textbf{84}, 074016 (2011)}.

\bibitem{Chen:2011pv}
  D.~Y.~Chen and X.~Liu,
  $Z_b(10610)$ and $Z_b(10650)$ structures produced by the initial single pion emission in the $\Upsilon(5S)$ decays,
  \href{http://dx.doi.org/10.1103/PhysRevD.84.094003}{Phys.\ Rev.\ D \textbf{84}, 094003 (2011)}.

\bibitem{Chen:2014ccr}
  D.~Y.~Chen, X.~Liu, and T.~Matsuki,
  Explaining the anomalous $\Upsilon(5S)\to \chi_{bJ}\omega$ decays through the hadronic loop effect,
  \href{http://dx.doi.org/10.1103/PhysRevD.90.034019}{Phys.\ Rev.\ D \textbf{90}, 034019 (2014)}.

\bibitem{Wang:2016qmz}
  B.~Wang, X.~Liu, and D.~Y.~Chen,
  Prediction of anomalous $\Upsilon(5S)\to\Upsilon(1^3D_J)\eta$ transitions,
  \href{http://dx.doi.org/10.1103/PhysRevD.94.094039}{Phys.\ Rev.\ D \textbf{94}, 094039 (2016)}.

\bibitem{Huang:2017kkg}
  Q.~Huang, B.~Wang, X.~Liu, D.~Y.~Chen, and T.~Matsuki,
  Exploring the $\Upsilon (6S)\rightarrow \chi _{bJ}\phi $ and $\Upsilon (6S)\rightarrow \chi _{bJ}\omega $ hidden-bottom hadronic transitions,
  \href{http://dx.doi.org/10.1140/epjc/s10052-017-4726-8}{Eur.\ Phys.\ J.\ C \textbf{77}, 165 (2017)}.

\bibitem{Huang:2018pmk}
  Q.~Huang, X.~Liu, and T.~Matsuki,
  Proposal of searching for the $\Upsilon(6S)$ hadronic decays into $\Upsilon(nS)$ plus $\eta^{(\prime)}$,
  \href{http://dx.doi.org/10.1103/PhysRevD.98.054008}{Phys.\ Rev.\ D \textbf{98}, 054008 (2018)}.

\bibitem{Huang:2018cco}
  Q.~Huang, H.~Xu, X.~Liu, and T.~Matsuki,
  Potential observation of the $\Upsilon(6S) \to \Upsilon(1^3D_J) \eta$ transitions at Belle II,
  \href{http://dx.doi.org/10.1103/PhysRevD.97.094018}{Phys.\ Rev.\ D \textbf{97}, 094018 (2018)}.

\bibitem{Belle:2018hjt}
  U.~Tamponi \textit{et al.} (Belle Collaboration),
  Inclusive study of bottomonium production in association with an $\eta $ meson in $e^+e^-$ annihilations near $\varUpsilon (5S)$,
  \href{http://dx.doi.org/10.1140/epjc/s10052-018-6086-4}{Eur.\ Phys.\ J.\ C \textbf{78}, 633 (2018)}
  
\bibitem{Cleven:2016qbn}
M.~Cleven and Q.~Zhao,
Cross section line shape of $e^+e^-\to\chi_{c0}\omega$ around the $Y(4260)$ mass region,
\href{http://dx.doi.org/10.1016/j.physletb.2017.02.041}{Phys.\ Lett.\ B \textbf{768}, 52-56 (2017).}
  
  
\bibitem{Khodjamirian:2011jp}
  A.~Khodjamirian, C.~Klein, T.~Mannel, and Y.~M.~Wang,
  Form Factors and Strong Couplings of Heavy Baryons from QCD Light-Cone Sum Rules,
  \href{http://dx.doi.org/10.1007/JHEP09(2011)106}{JHEP \textbf{09}, 106 (2011)}.

\bibitem{Cheng:2004ru}
  H.~Y.~Cheng, C.~K.~Chua and A.~Soni,
  Final state interactions in hadronic B decays,
  \href{http://dx.doi.org/10.1103/PhysRevD.71.014030}{Phys.\ Rev.\ D\ \textbf{71}, 014030 (2005)}

  \bibitem{BESIII:2019gjc}
  M.~Ablikim \textit{et al.} (BESIII Collaboration),
  Cross section measurements of $e^+ e^-\to\omega\chi_{c0}$ from $\sqrt{s}=$ 4.178 to 4.278 GeV,
  \href{http://dx.doi.org/10.1103/PhysRevD.99.091103}{Phys.\ Rev.\ D\ \textbf{99}, 091103 (2019)}

\bibitem{Dong:1994zj}
  Y.~B.~Dong, Y.~W.~Yu, Z.~Y.~Zhang and P.~N.~Shen,
  Leptonic decay of charmonium,
  \href{http://dx.doi.org/10.1103/PhysRevD.49.1642}{Phys.\ Rev.\ D\ \textbf{49}, 1642 (1994)}



\bibitem{Ernst:1960zza}
  F.~J.~Ernst, R.~G.~Sachs and K.~C.~Wali,
  Electromagnetic form factors of the nucleon,
  \href{http://dx.doi.org/10.1103/PhysRev.119.1105}{Phys.\ Rev.\ \textbf{119}, 1105 (1960)}.

\bibitem{BESIII:2021cvv}
  M.~Ablikim \textit{et al.} [BESIII],
  Measurement of $\Lambda$ baryon polarization in $e^+e^-\rightarrow\Lambda\bar\Lambda$ at $\sqrt{s} = 3.773$ GeV,
  \href{http://dx.doi.org/10.1103/PhysRevD.105.L011101}{Phys.\ Rev.\ D \textbf{105}, L011101 (2022)}.


\end{thebibliography}
\end{document}